\documentclass[preprint]{aastex63}
\usepackage{multirow}
\usepackage{booktabs}

\submitjournal{ApJL}

\shorttitle{Comparison of ACR and GCR oxygen at 1 au}
\shortauthors{Fu et al.}
\graphicspath{{./}{figures/}}
\begin{document}

\title{Comparison of Anomalous and Galactic Cosmic Ray Oxygen at 1 au during 1997--2020}

\author[0000-0003-4245-3107]{Shuai Fu}
\altaffiliation{These authors contributed equally.}
\affiliation{State Key Laboratory of Lunar and Planetary Sciences, Macau University of Science and Technology, Taipa 999078, Macau, PR China}
\affiliation{CNSA Macau Center for Space Exploration and Science, Taipa 999078, Macau, PR China}

\author[0000-0002-4299-0490]{Lingling Zhao}
\affiliation{Center for Space Plasma and Aeronomic Research (CSPAR), University of Alabama in Huntsville, Huntsville, AL 35805, USA}

\author[0000-0002-4306-5213]{Xiaoping Zhang}
\altaffiliation{These authors contributed equally.}
\affiliation{State Key Laboratory of Lunar and Planetary Sciences, Macau University of Science and Technology, Taipa 999078, Macau, PR China}
\affiliation{CNSA Macau Center for Space Exploration and Science, Taipa 999078, Macau, PR China}

\author[0000-0002-1066-2273]{Pengwei Luo}
\affiliation{State Key Laboratory of Lunar and Planetary Sciences, Macau University of Science and Technology, Taipa 999078, Macau, PR China}
\affiliation{CNSA Macau Center for Space Exploration and Science, Taipa 999078, Macau, PR China}

\author[0000-0002-5565-8382]{Yong Li}
\affiliation{State Key Laboratory of Lunar and Planetary Sciences, Macau University of Science and Technology, Taipa 999078, Macau, PR China}
\affiliation{CNSA Macau Center for Space Exploration and Science, Taipa 999078, Macau, PR China}

\correspondingauthor{Lingling Zhao}
\email{lz0009@uah.edu}

\begin{abstract}
Using quiet-time measurements of element oxygen within the energy range 7.3--237.9 MeV nuc$^{-1}$ from the ACE spacecraft at 1 au, we compare the energy spectra and intensities of anomalous and Galactic cosmic rays (ACRs and GCRs, respectively) during 1997--2020. Our analysis shows that the transition from ACR-dominated spectrum to GCR-dominated spectrum occurs at energies $\sim$15 to $\sim$35 MeV nuc$^{-1}$, and the transition energy $E_t$ is found to be well anticorrelated with varying solar activity. This is the first study of ACR-GCR transition energy dependence on the solar cycle variation. At energies below $E_t$, the index of the power-law ACR-dominated spectrum ($\gamma_1$) ranges from -2.0 to -0.5, whereas the GCR-dominated spectrum has a power-law index ($\gamma_2$) changing from 0.3 to 0.8 at energies ranging from $E_t$ to 237.9 MeV nuc$^{-1}$. Both $\gamma_1$ and $\gamma_2$ are positively correlated with solar activity. In addition, during the solar cycle 24/25 minimum period, the peak GCR intensity observed by ACE spacecraft is about 8\% above its 2009 value, setting a new record since the space age, while the peak ACR intensity is almost similar to that of the previous two solar cycles with the same pattern of solar magnetic polarity, indicating a different modulation mechanism between ACRs and GCRs.
\end{abstract}

\section{Introduction}
\label{sect1}
Energetic particles commonly observed in the heliosphere are composed of solar energetic particles (SEPs, also known as solar cosmic rays, with energies of a few keV up to several GeV, \citealp{fu19}), Galactic cosmic rays (GCRs, with energies of tens of MeV nuc$^{-1}$ up to $10^{14}$ MeV nuc$^{-1}$, \citealp{heilbronn20}), and anomalous cosmic rays (ACRs, with energies of $\sim$5--50 MeV nuc$^{-1}$, \citealp{Rankin21}). These high-energy charged particles produce severe radiation in space and pose a great threat to spacecrafts and astronauts carrying out deep space missions.

SEPs originate from the Sun and are frequently associated with solar flares and coronal mass ejections (CMEs). CME-driven shock waves are an efficient accelerator for solar particles (e.g., \citealt{fu19}). GCRs are generally believed to come from outside the solar system within the Milky Way, but it does not rule out the possibility of other galaxy sources (e.g., \citealp{Ackermann13,Aab17}). There is rather convincing evidence that supernova remnant shock waves initiated by supernova explosions are the prime GCR acceleration source (e.g., \citealp{Aharonian07}). ACR component was first reported to be an anomalous enhancement in the low energy end of the helium spectrum during the solar quiet periods in 1972 (\citealp{Garcia-Munoz73}). A prevailing theory for explaining the source of ACRs was given by \citet{Fisk74}, suggesting that these particles might originate from interstellar neutral atoms that drift into the solar system; they become singly ionized by solar radiation and/or charge exchange with the solar wind, and then are carried out to the outer heliosphere by the solar wind as ``pickup'' ions (PUIs) and get accelerated up to energies of tens to hundred MeV, presumably at the heliospheric termination shock (HTS) by diffusive shock acceleration (e.g., \citealp{pesses81,Giacalone12}). These accelerated particles suffer solar modulation during their propagation in the heliosphere and are finally observed as ACRs. Therefore, it is believed that ACRs propagating from their acceleration site back towards Earth can transmit information from the HTS to the inner heliosphere (e.g., \citealp{Fichtner01}). Prior to the Voyager crossings of the HTS, the interstellar PUIs experience diffusive shock acceleration at the HTS was the long-held paradigm for explaining ACR generation, and the HTS is widely considered as the source location of ACRs \citep{Fisk74,pesses81}. Surprisingly, the measured ACR intensity profiles from Voyagers 1 \& 2 did not peak at the HTS as previously expected, but peaked within $\sim$1 au behind the HTS, indicating that the source of ACRs may not be in the region of the HTS crossed by Voyager spacecraft (e.g., \citealp{Stone05,Stone08,cummings13,zhao19}). These findings contradict the conventional ACRs generation paradigm, and directly lead to the proposal of many alternative theories for ACR generation. \citet{McComas06} firstly suggested that the blunt shape (nose-to-tail asymmetry) of the shock causes the nonlocal diffusive acceleration at the HTS, providing a natural explanation for the high-energy ACR intensity not peaking at the shock. Later, \citet{kota2008} and \citet{Schwadron08} presented numerical simulations modeling the acceleration of ACRs at a blunt HTS, and their predictions are consistent with Voyager spacecraft observations, especially in the inner heliosheath. Besides, there are some other ACR acceleration mechanisms, including magnetic reconnection in the sectored heliosheath (e.g., \citealp{Drake10,Zank15,zhao19}), acceleration by random plasma compressions (e.g., \citealp{Fisk09}), second-order statistical acceleration  (e.g., \citealp{Zhang06,Strauss10}). 

Previous observations show that there are significant differences in elemental compositions and charge states between ACRs and GCRs. GCRs are generally thought to originate from supernovae explosions, from kiloparsecs away (e.g., \citealp{Ackermann13}), and they are composed of fully stripped atomic nuclei (98\%) and electrons (2\%), and the majority of the nucleonic component is hydrogen nuclei ($\sim$87\%), helium nuclei ($\sim$12\%), and heavier nuclei ($\sim$1\%) \citep{Simpson83}. ACRs ultimately originate from interstellar neutrals, which requires that the most abundant ACR species should be the elements with high first ionization potentials (FIPs). The major constituents of ACRs in the outer heliosphere are found to be H, He, N, O, Ne, and Ar, and all of them except H are measured in the inner heliosphere (e.g., \citealp{Hovestadt73,McDonald74, Cummings02}). For elements with FIPs lower than that of H, most of them have been ionized in the local interstellar medium and are deflected by the interplanetary magnetic field as they approach the heliosphere (e.g., \citealp{Cummings02,Cummings07}). Unlike GCRs which are always fully ionized, low-energy ACRs are mostly singly charged because the acceleration occurs faster than additional electron stripping, but they can be multiply charged at higher energies (above $\sim$20 MeV nuc$^{-1}$) as a consequence of further charge-stripping in the acceleration process (e.g., \citealp{klecker1995,Cummings07}). The different charge states of ACRs and GCRs can affect their propagation and lead to significantly different modulation behaviors. In general, particles with higher charge states cause the drift and diffusion to be slower, which prevents them from penetrating into the inner heliosphere readily (e.g., \citealp{Jokipii96}). It is reported that the observed GCR radial intensity gradients are much smaller than that of ACRs, which is probably due to the difference between their charge states (e.g., \citealp{Cummings87,Leske05}). Based on the recent observations from Parker Solar Probe, \citet{Rankin21} have studied the radial gradient of ACRs in to $\sim$36 solar radii  (0.166 au) and  reported that the radial gradients of ACR helium at energies of $\sim$4 to $\sim$45 MeV nuc$^{-1}$ are larger than previous observations made further out in the heliosphere.

Cosmic-ray (CR) intensities are sensitive to changes in solar activity, the so-called ``solar modulation''. Convection, diffusion, drifts, and adiabatic deceleration are the four major processes of solar modulation. Typically, the enhanced diffusion associated with the turbulent magnetic field of solar maximum may be the dominant factor in affecting the transport of CRs through the heliosphere, but drift effects are relatively more important around solar minimum (e.g., \citealp{zhao15}). Drift effects are related not only to the gradient and curvature of the interplanetary magnetic field, but also to the wavy heliospheric current sheet (HCS). The drift of charged particle differs between different solar magnetic cycles, generally speaking, positively charged particles predominantly undergo an inward drift along the HCS when the solar magnetic field at the north pole points inward (negative polarity, $A<0$), whereas positively charged particles drift inward from the heliospheric polars to the solar equator and then drift outward along the HCS when the solar magnetic field at the north pole points outward (positive polarity, $A>0$) (e.g., \citealp{jokipii81}). CR particles arrive at the Earth more easily when their route of access is through the polar regions than when they reach the Earth along the HCS \citep{Mavromichalaki07}. There is a tight negative correlation between HCS tilt angle and CR intensity, for both solar magnetic polarities \citep{Belov2000}. However, the time lag of CRs to solar activity behaves a positive dependence on HCS tilt angle, attributing to the fact that CRs must experience a longer access route to reach the Earth when the HCS is tilted and disturbed \citep{Mavromichalaki07}.

During solar quiet periods, the energy spectra of ACRs and GCRs overlap at low energies, especially at energies below $\sim$50 MeV nuc$^{-1}$ (e.g., \citealp{Simnett17}). Based on the 1974--1978 ISEE-3 spacecraft measurements, \citet{Mewaldt84} claimed that ACR component dominates the quiet-time spectra of N and O nuclei within 5--30 MeV nuc$^{-1}$. With measurements taken from the SAMPEX and GEOTAIL spacecrafts over the 1992--1993 time period, \citet{Mewaldt93} and \citet{Hasebe94} reported that the spectra of elements N, O and Ne below $\sim$30 MeV nuc$^{-1}$ have prominent enhancements in flux above the quiet-time GCR spectra, which is the typical feature of ACR component. \citet{leske13} studied the ACE measurements during the solar minima 1997/1998 and 2009 and also found an apparent turning point at the energy $\sim$30 MeV nuc$^{-1}$ in the spectra of N, O and Ne. ACR component is regarded as the major energetic particle species at energies lower than 50 MeV nuc$^{-1}$ in the outer heliosphere \citep{Webber05}.

In this paper, we aim to conduct a study on the long-term CR variations based on in-situ measurements at 1 au. The paper is organized as follows. Section \ref{sect2} provides an overview of the dataset and the method used for determining the transition location from ACR-dominated spectrum to GCR-dominated spectrum\footnote{
In this paper, ``ACR-dominated spectrum'' and ``GCR-dominated spectrum'' generally refer to the observations in which the ACR and GCR components are expected to dominate, respectively; ``ACR-GCR transition'' refers to the transition from ACR-dominated spectrum to GCR-dominated spectrum.}. The analysis of both ACR and GCR spectra and intensities are presented in Section \ref{sect3}. The discussions and conclusions are drawn in Section \ref{sect4}.

\section{DATASET AND METHOD}
\label{sect2}
\subsection{ACE observations}

The fluxes of ACRs and GCRs at 1 au are measured by the Solar Isotope Spectrometer (SIS) and the Cosmic Ray Isotope Spectrometer (CRIS) instruments aboard the Advanced Composition Explorer (ACE) spacecraft, respectively  \citep{Stone98a}. The two instruments have continuously recorded highly accurate fluxes of multiple CR nuclei ($2 \le Z \le 28$) since 1997, spanning solar cycles 23 and 24.

\begin{figure}[ht!]
\epsscale{1.05}
\plotone{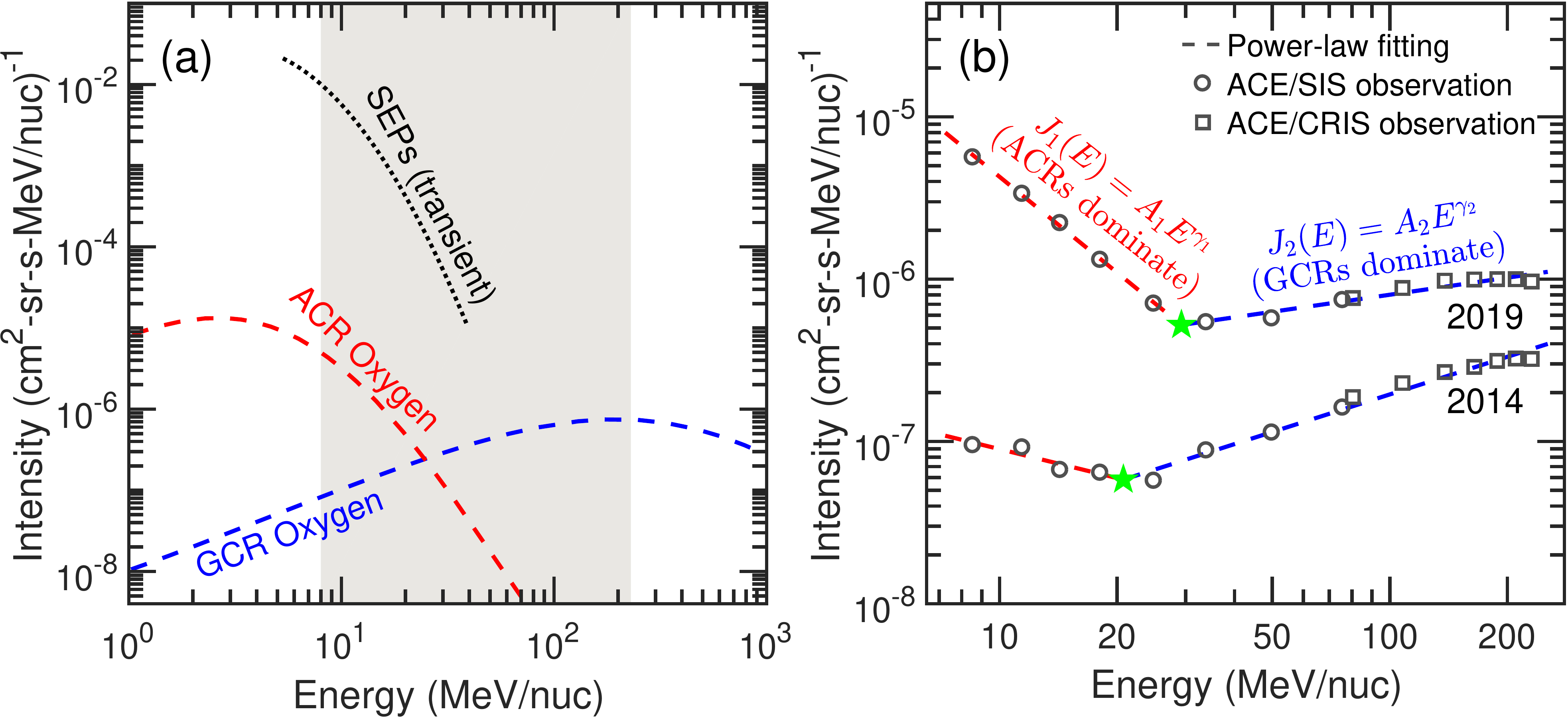}
   \caption{(a) Diagrammatic sketch of the energy spectra measured for SEP, ACR and GCR oxygen, reproduced from \citet{Stone98b}. The shaded region indicates the studied energy range (7.3--237.9 MeV nuc$^{-1}$). (b) The observed energy spectra for oxygen in the years 2014 and 2019. The circles and squares are the measurements from SIS and CRIS, respectively. The red and blue dashed lines are the power-law fits to the ACR-dominated and GCR-dominated data points, respectively. The green pentagram denotes the optimal transition location ($E_t$).}  \label{Fig1}
\end{figure}

Here, we use the re-evaluated level-2 daily averaged oxygen flux data (in units of $particles/(cm^2 \cdot sr \cdot s \cdot MeV \ nuc^{-1}$)) provided by the ACE Science Center (ASC) during the period from 1997 August to 2020 December. Figure \ref{Fig1}(a) illustrates a schematic diagram of the energy spectra of the major energetic particle populations that can be observed by ACE/SIS and CRIS instruments. The shaded area represents the energy range studied in this work, which is from $\sim$7.3 to 237.9 MeV nuc$^{-1}$ and contains a mixture of SEP, ACR, and GCR components. Given that the measured CR fluxes suffer severe contamination from SEP events (\citealp{leske13}), especially around solar maximum, we use the solar activity flag provided by ACE/SIS data to remove the SEP component. After excluding the contribution from SEP events and combining the GCR measurement from ACE/CRIS instrument, we can obtain a complete CR data set within the energy range 7.3--237.9 MeV nuc$^{-1}$. Then, we compute the 27-day Bartels Rotation average fluxes to eliminate the effects of sporadic and short-term ($< $27 day) disturbances originating at the Sun, such as solar flares and CMEs, and mainly focus on longer-scale periodic variations. Note that the interval is entirely discarded if it includes good data less than half a Bartels rotation ($\sim$13.5 days) to ensure that our results are statistically reliable.

\subsection{Determination of ACR-GCR transition location}

Assuming CRs are accelerated via shock waves, the energy spectra of ACRs and GCRs are expected to be a power law. As shown in Figure \ref{Fig1}(b), the ACE observed spectra of ACR-dominated and GCR-dominated parts can be individually fitted by one power-law function with the form of $J \propto E^ \gamma$, where $J$ is the differential flux, and $E$ is the kinetic energy.  Therefore, if the ACR-GCR transition occurs at an energy of $E_t$, the energy spectra for oxygen within the energy range 7.3--237.9 MeV nuc$^{-1}$ are described by,
\begin{equation}
\left\{ \begin{array}{ccc}
J_1(E)=A_1\cdot E^{\gamma_1},   \ for\   7.3 \leq E\leq E_t; \\
J_2(E)=A_2\cdot E^{\gamma_2},   \ for\   E_t \leq E\leq 237.9,
\end{array} \right.
\label{eq1}
\end{equation}
where $A_i$ ($i=1, 2$) is a constant and its magnitude essentially equals to the extrapolated ACR-dominated or GCR-dominated flux at an energy of 1.0 MeV nuc$^{-1}$, $\gamma_i$ is the spectral index (with the subscript ``1'' for the ACR-dominated spectrum, and the subscript ``2'' for the GCR-dominated spectrum), and $E_t$ is the transition energy.

The algorithm for determining the optimal transition/breakpoint location is described below. First of all, after logarithmic transformation of the original data, the problem is essentially a piecewise linear fitting problem with an unknown breakpoint. The determination of the optimal breakpoint is an optimization problem, and is formulated to find a proper breakpoint that minimizes the overall residual sum of squares ($RSS=\sum\limits_{}^{}(J_{fit}-J_{obs})^2$, where $J_{fit}$ and $J_{obs}$ denote the fitted and observed values, respectively). To do so,  we need to traverse a set of possible breakpoint locations within the energy range 7.3--237.9 MeV nuc$^{-1}$. For a given location, a least squares fit is performed which solves for the regression parameters through minimizing the respective \emph{RSS}s, and the overall \emph{RSS} is also computed in this case. After traversing all possible breakpoint locations, we derive a set of overall \emph{RSS}s, and the optimal breakpoint location should minimize $RSS$s. The Python package \emph{pwlf} is used here to solve piecewise function and determine the optimal breakpoint location (\citealp{jekel2019pwlf}).

In Figure \ref{Fig1}(b), we show the observed CR oxygen spectra in year 2014 and year 2019 to represent solar maximum and solar minimum, respectively. It is clear that the spectrum in 2019 is much higher than that in 2014. Compared with 2014, the flux increase of low energy particles in both ACR and GCR components is more than that of high energy particles in 2019. It indicates that the solar modulation effect is more significant for low energy particles, which applies for both ACR and GCR components. We also note that the spectral transition energy marked by the green pentagram has a clear increase in 2019 compared to 2014. Furthermore, the ACR-dominated spectrum in 2014 is flatter than that in 2019 and both have a negative spectral index. The GCR-dominated spectrum in 2014 with a positive spectral index is steeper compared to that in 2019.

\section{Results and analysis}
\label{sect3}
\subsection{Comparison of ACR-dominated and GCR-dominated spectra}
\label{subsect3.1}

\begin{figure}[ht!]
\epsscale{0.8}
\plotone{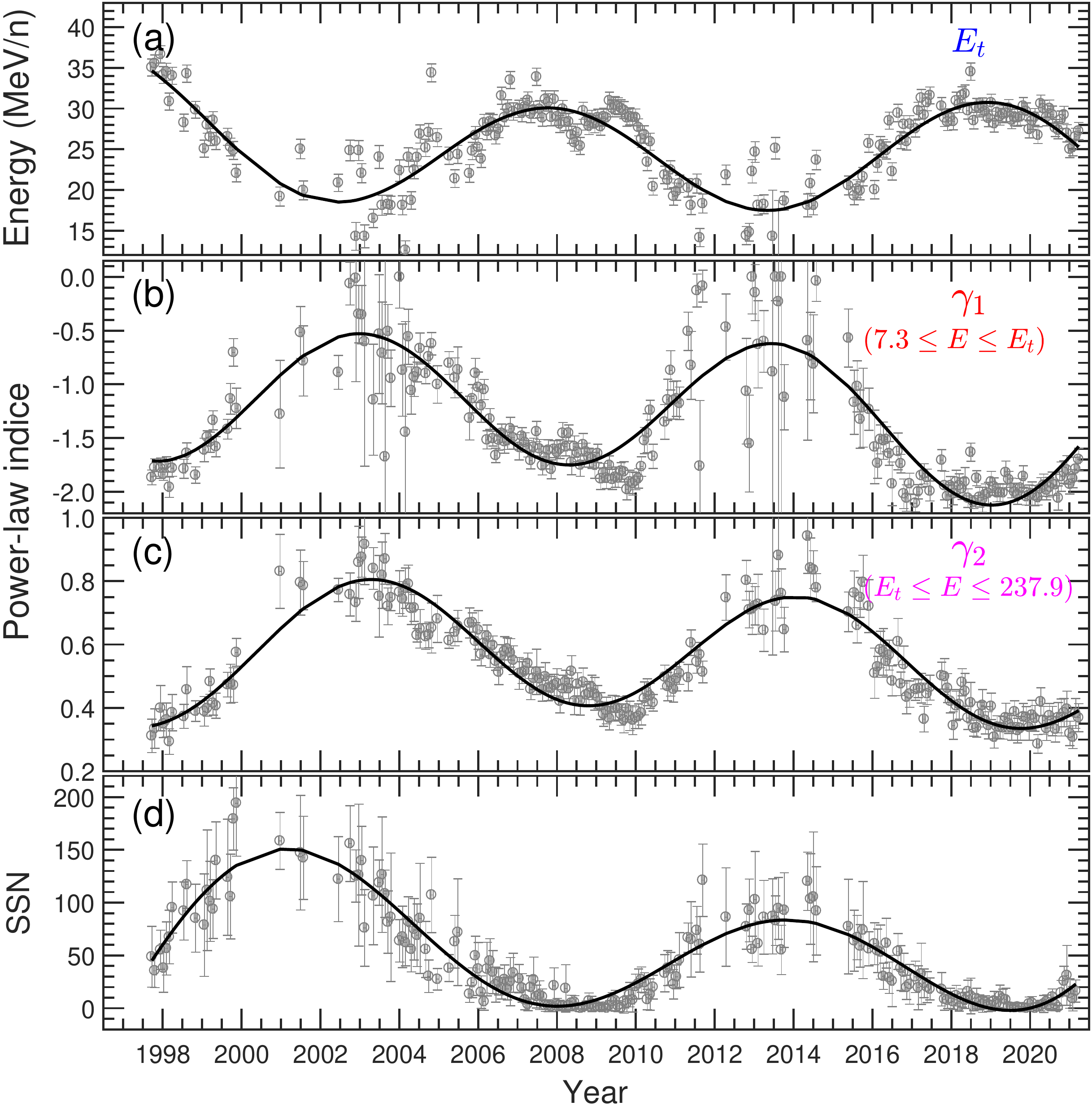}
   \caption{From top to bottom: ACR-GCR transition energy ($E_t$), spectral slope of ACR-dominated spectrum ($\gamma_1$), spectral slope of GCR-dominated spectrum ($\gamma_2$), 27-day averaged sunspot number. Calculation errors are also shown for each parameters. The thick black lines are the fits to the scattering points using the sum of sine functions. See text for details. \label{Fig2}}
\end{figure}

Using the method mentioned in Section \ref{sect2}, we fit the observed energy spectrum in every 27-day interval. Figures \ref{Fig2}(a)--(c) show the best-fit values of parameters $E_t$, $\gamma_1$, and $\gamma_2$, respectively, and Figure \ref{Fig2}(d) are the 27-day averaged sunspot numbers (SSNs) from 1997 to 2020. We note that the error estimation in Figures \ref{Fig2}(a)--(c) represents a 95\% confidence interval for each parameter determined from the fitting process, and in Figure \ref{Fig2}(d) the error bar is calculated by the standard deviation in calculating the 27-day averaged SSN. In addition, the sum of sine function is used to fit these scatter points with obvious periodic variations, which is described by
$y_j = \sum\limits_{i=1}^{s} a_i \cdot sin(b_i \cdot x_j + c_i), \ i=1,2,...,s, \ j=1,2,...,n,$
where $s$ is the number of terms in the series (generally with $1 \le s \le 8$, here we use $s=3$), $n$ is the number of scatter points, $a_i$, $b_i$ and $c_i$ are the amplitude, the frequency, and the horizontal phase constant of each sine wave term, respectively. Nonlinear least squares is used to fit the function to our data. Clearly, all the spectra parameters ($E_t$, $\gamma_1$, and $\gamma_2$) vary dramatically with the solar activity cycle.

Figure \ref{Fig2}(a) shows that the determined ACR-GCR transition energy $E_t$ changes from $\sim$15 to $\sim$35 MeV nuc$^{-1}$, which is in agreement with earlier studies using measurements from ISEE-3, SAMPEX, GEOTAIL and ACE spacecrafts (e.g., \citealp{Mewaldt84,Mewaldt93,Hasebe94,leske13}). Besides, $E_t$ shows a clear inverse correlation with SSN variation, reaching peaks around solar minima and valleys near solar maxima. To the best of our knowledge, this is the first time that a transition from ACR to GCR in two cycles has been reported. We speculate that such an anticorrelation between $E_t$ and SSN is essentially related to the difference between solar modulation of ACRs and GCRs during their respective propagation within the heliosphere. ACRs are known to dominate the low energy end of the CRs spectra and therefore are more susceptible to the solar variations than GCRs. As shown in Figures \ref{Fig4}(b) and \ref{Fig4}(c), the intensities of ACR oxygen are enhanced by more than two orders of magnitude from solar maximum to solar minimum, while the simultaneous GCR intensities vary only within one order of magnitude. Therefore, during solar quiet periods, the intensity of ACRs increase significantly more than that of GCRs, leading to an extended ACR-dominated range and higher transition energy. On the other hand, weak solar activity also weakens the solar modulation of GCRs propagation. For the same energy seed particles in the outer heliosphere, they can undergo less solar modulation and reach the Earth at higher energies. Instead, GCR particle needs to experience stronger solar modulation during solar active periods, which results in a lower energy of GCR observed at 1 au and a natural decline in the ACR-GCR transition energy.

Figures \ref{Fig2}(b) and \ref{Fig2}(c) show the fitted power-law indexes of ACR- and GCR-dominated spectra, respectively. We can see that ACR-dominated spectral indexes, $\gamma_1$, are negative and mainly vary from -2.0 to -0.5, for the energy range 7.3 MeV nuc$^{-1}$ to $E_t$; GCR-dominated spectra within the energy range from $E_t$ to 237.9 MeV nuc$^{-1}$ are positively indexed with $\gamma_2$ ranging from 0.3 to 0.8. Both $\gamma_1$ and $\gamma_2$ are positively correlated with varying solar activity. To be specific, ACR-dominated spectrum experiences a hardening at solar maximum and a softening at solar minimum, whereas GCR-dominated spectrum becomes softer at solar maximum and gets harder at solar minimum. This variation can be understood from the different modulation between low-energy and high-energy particles. It is commonly known that low-energy particles are more sensitive to varying solar activity compared to high-energy ones. At solar minimum, the flux enhancement of low-energy particles is larger than that of high-energy particles, in both ACR and GCR components, causing the ACR-dominated spectral softening while the GCR-dominated spectral hardening (below 250 MeV nuc$^{-1}$). On the other hand, at solar maximum, the decrease of low-energy particles is more significant than that of high-energy particles, causing the ACR-dominated spectral hardening while the GCR-dominated spectral softening. This is clearly illustrated in Figure \ref{Fig1}(b).

To quantify the correlation between CR spectral parameters ($E_t$, $\gamma_1$, $\gamma_2$) and SSN, the Pearson correlation coefficient is calculated,
\begin{equation}
r=\frac{\sum\limits_{i=1}^{n}(x_{i}-\bar{x})(y_{i}-\bar{y})} {\sqrt{\sum\limits_{i=1}^{n}(x_{i}-\bar{x})^2}\sqrt{\sum\limits_{i=1}^{n}(y_{i}-\bar{y})^2}},
\label{eq3}
\end{equation}
where $x$ and $y$ are a pair of variables. The calculated correlation coefficients between $E_t$, $\gamma_1$, $\gamma_2$ and SSN are -0.61, 0.70, 0.74, respectively, indicating a statistically significant positive correlation between CR spectral slope and SSN and a negative correlation between transition energy and SSN.

\begin{figure}[ht!]
\epsscale{0.52}
\plotone{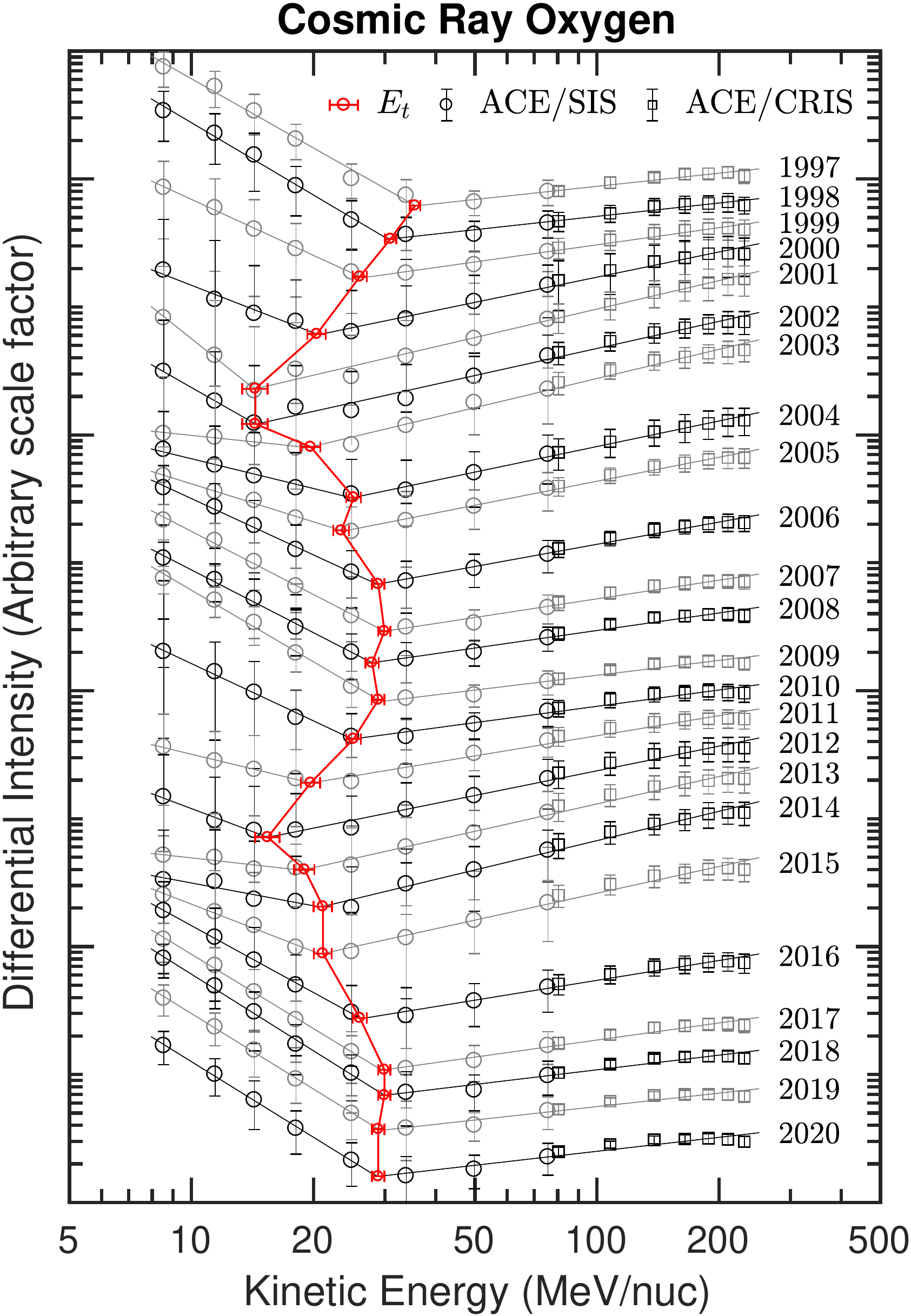}
   \caption{Annual energy spectrum of oxygen. The solid curves are the piecewise power-law fits to the ACE measured data, and the red circles denote the ACR-GCR transition energy. The uncertainty on $E_t$ represents a 95\% confidence interval determined by piecewise fitting; the error bar of ACE data is calculated by the standard error in calculating the yearly averaged CR intensities.} Arbitrary scale factors are applied in the spectra for the presentation purpose.  \label{Fig3}
\end{figure}

Figure \ref{Fig3} displays the yearly averaged CR spectra of oxygen and the corresponding determined ACR-GCR transition energy. Clearly, the power-law shape of both ACR and GCR spectra changes greatly from year to year, with flat or steep features, which is also reflected in Figure \ref{Fig2}. Moreover, the annual ACR-GCR transition energy also exhibits a notable solar cycle dependence as marked by the red dashed line. Note that the magnitude of the flux, which can be briefly seen in Figure \ref{Fig1}(b), is not comparable due to the application of arbitrary scale factors in the spectra.

\subsection{Comparison of ACR and GCR intensities}
\begin{figure}[ht!]
\epsscale{0.8}
\plotone{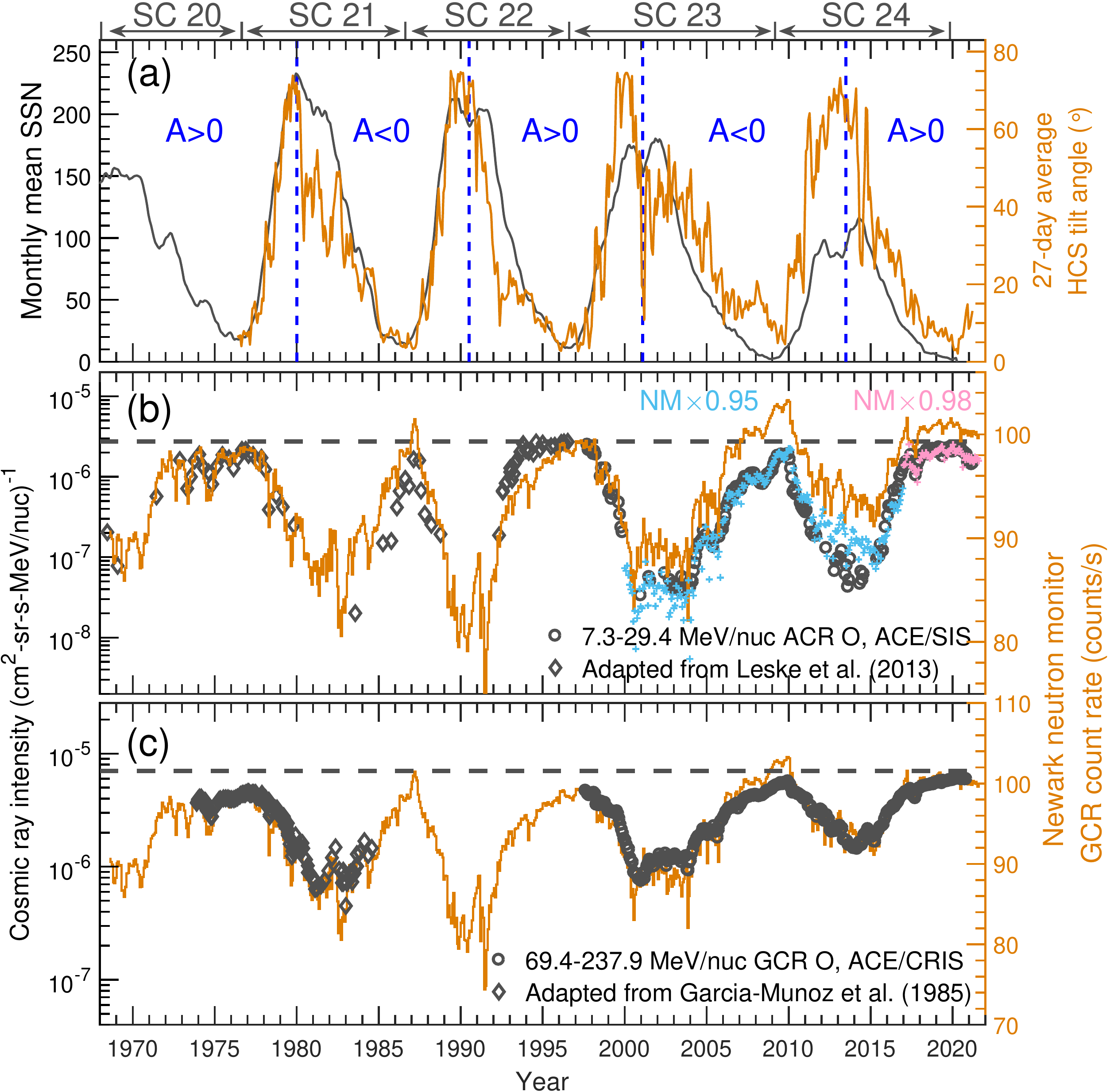}
   \caption{(a) Monthly smoothed sunspot number, compared with the ``radial model'' computed HCS tilt angle. (b) 27-day quiet-time average intensities of ACR oxygen in space (circles are from ACE/SIS, and diamonds are adapted from \citet{leske13} at energies 8--27 MeV nuc$^{-1}$). (c) 27-day average intensities of GCR oxygen in space (circles are from ACE/CRIS, and diamonds are adapted from IMP-8 at energies 53--211 MeV nuc$^{-1}$ \citep{Garcia-Munoz85}. GCR count rates from Newark station are plotted in panels (b)--(c) as a reference. The horizontal dashed lines in panels (b) and (c) mark the peak intensity of ACRs and GCRs during the 2019--2020 solar minimum, respectively.  \label{Fig4}}
\end{figure}

The maximum amplitude of the solar cycle has been decreasing since the 1980s, and the recently complete solar cycle 24 is recorded to be the weakest one  in the space age \citep{fu21}, as shown in Figure \ref{Fig4}(a). The HCS tilt angle computed from the ``radial model" is $\sim$2.1$^\circ$ in 2020 April, the lowest value since the year 1976. This extremely flat HCS makes it easier for CR particles to arrive at the Earth than the previous $A > 0$ solar minimum, as a consequence of the shortened propagation path. Additionally, ACR intensities observed near the Earth are believed to be more sensitive to the HCS tilt angle than GCRs due to the latitude dependence of ACR source intensity (e.g., \citealp{leske13,Simnett17}).

Figures \ref{Fig4}(b) and \ref{Fig4}(c) show the extended observations of ACR and GCR oxygen in space, respectively. The GCR count rates measured by the ground-based Newark neutron monitor station (cutoff rigidity 2.40 GV) are also plotted as a reference.

The most striking feature of Figure \ref{Fig4}(b) is that the observed ACR intensities\footnote{Of note here is that for comparison with the historical data adapted from \citet{leske13}, we select the energy range 7.3--29.4 MeV nuc$^{-1}$ as ACR oxygen which includes a portion of GCRs according to Figure \ref{Fig2}(a), but it does not affect our analysis since we mainly focus on the solar minimum period when $E_t$ is relatively large.} are inversely correlated with varying solar activity, and the peak value of ACR intensity behaves a clear solar magnetic polarity dependence as a result of particle drift effects. Specifically, the maximum ACR intensity during the solar minimum $P_{24/25}$ is almost similar to (or slightly lower than) that during the solar minina $P_{20/21}$ and $P_{22/23}$ ($A>0$ cycles), whereas the peak ACR intensity in late 2009 is close to that in 1987 ($A<0$ cycles). Furthermore, ACRs can reach higher peak intensities in $A>0$ solar minima than in $A<0$ solar minima (e.g., $\sim$23\% higher in 2019 than in 2009), ascribing to their different propagation paths in the heliosphere corresponding to different polarities (as mentioned in the Introduction section).

In Figure \ref{Fig4}(b), the ACR time-intensity profile follows the GCR count rate well in the 1970s and 1997--2000, but they deviate from each other since 2000 when the solar magnetic polarity is reversed. From 2000--2016, the good consistency between the ACR intensities and the GCR count rates can be reproduced by reducing the GCR count rates to $\sim$0.95 times (cyan symbols). The artificially modified GCR count rates deviate from the ACR profile again around 2017, but the two profiles can be adjusted to be consistent by reducing the original GCR count rates to a factor of $\sim$0.98 (pink symbols). The overall GCR scale factor has changed from $\sim$0.95 to $\sim$0.98 over the past 20 years, which may relate to the weakening solar cycles.

Figure \ref{Fig4}(c) shows the observed GCRs both in space and on the ground. The GCR time-intensity profile tracks the GCR count rates fairly well for the periods when spacecraft measurements are available, and the maximum GCR intensities also exhibit an obvious polarity dependence, with alternating peaked ($A < 0$) and flat-topped ($A > 0$) shapes during successive solar minima. The maximum GCR intensities in space set a record during the solar minimum $P_{24/25}$, exceeding those recorded in 1997 and 2009 by $\sim$30\% and $\sim$8\%, respectively.

\begin{figure}[ht!]
\epsscale{1.0}
\plotone{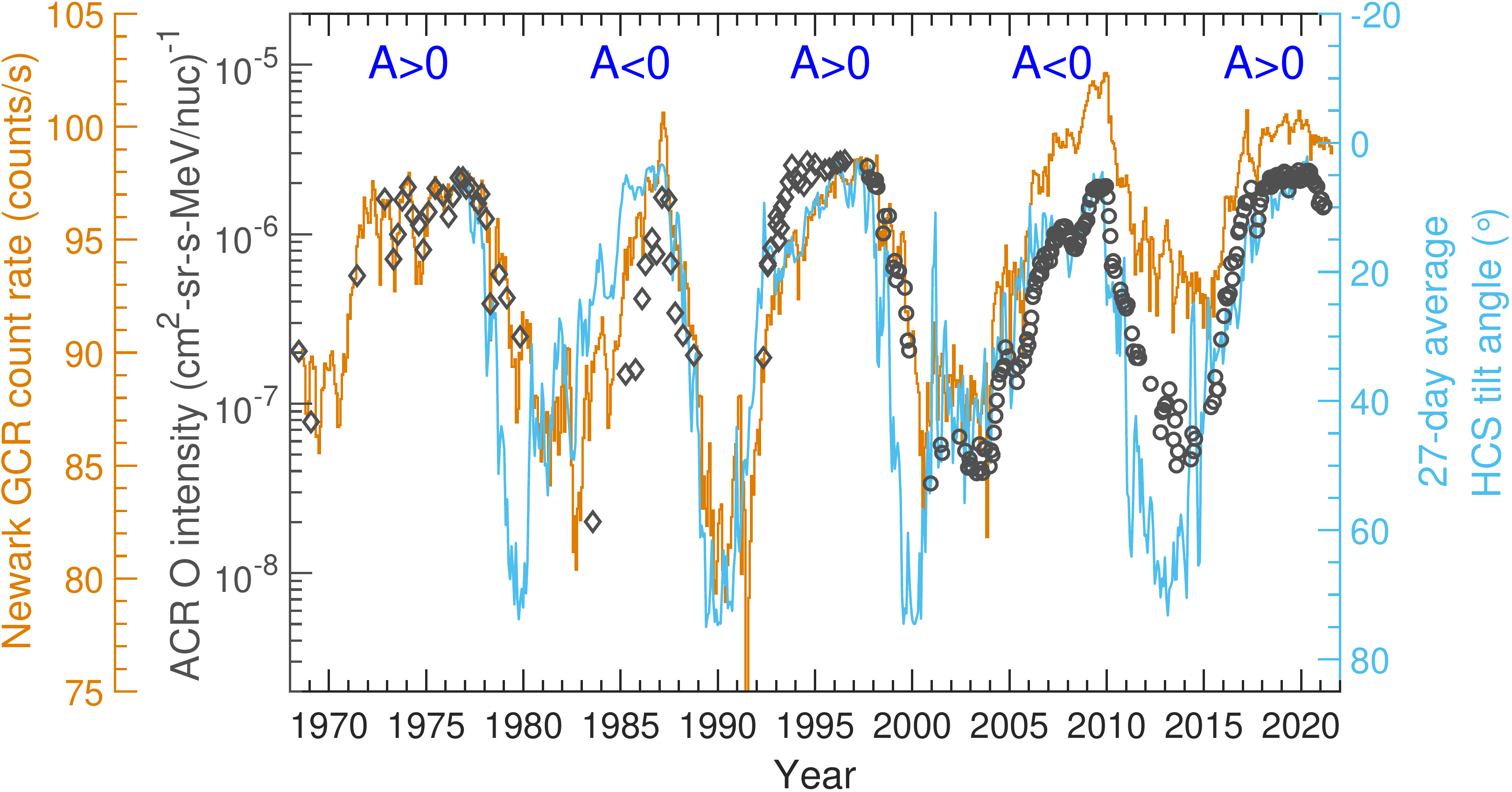}
   \caption{Comparision of the NM count rates (orange curve; \emph{first left axis}), the ACR oxygen intensities in space (black data points; \emph{second left axis}) with the HCS tilt angle (cyan curve; \emph{right axis}). An inverted scale is applied in the \emph{right axis} for more intuitive comparison.}   \label{Fig5}
\end{figure}

The unusual quiet solar cycle 24 should result in a significant decrease in CR modulation and a corresponding significant increase in CR intensity (including both ACR and GCR components). However, the observed ACR intensity in 2019--2020 is roughly the same as that observed in the previous two $A>0$ solar minima, although it is elevated compared to that in 2009. One possibility is that the ACR intensities observed at 1 au are being inhibited by some factors, such as a decrease of ACR seeds, a reduction of ACR acceleration efficiency, or a greater sensitivity of ACRs to the HCS tilt angle  \citep{leske18}. Voyager spacecrafts have entered interstellar space, and their observations may provide more answers to this question in the future.
By comparing ACR and GCR time-intensity profiles with the solar cycle evolution of the HCS tilt angle profile, as shown in Figure \ref{Fig5}, we find that the ACR time-intensity profile measured by ACE/SIS tracks the inverted HCS tilt angle closely in most of the observational periods since the year 1997 (except around solar maximum), but the GCR time-intensity profile deviates from the HCS tilt angle solar cycle variation completely after the year 2006. It demonstrates that ACRs are more sensitive to the HCS tilt angle than GCRs, and suggests that current sheet drift may play an important role in ACR modulation (e.g., \citealp{leske13}).

\section{Discussion and Conclusion}
\label{sect4}
In this paper, the differential energy spectra and intensities of ACR and GCR oxygen within 7.3--237.9 MeV nuc$^{-1}$ are investigated using measurements from ACE spacecraft at 1 au during the solar cycles 23 and 24. The paper is concluded as follows.

(1) The transition from ACR-dominated spectrum to GCR-dominated spectrum occurs at energies $\sim$15 to $\sim$35 MeV nuc$^{-1}$, which is  consistent with previous findings. Moreover, the transition energy $E_t$ shows a significant anti-correlation with varying solar activity. This is the first report of the change in the ACR-GCR transition relative to the solar cycle variation, benefiting from ACE/CRIS and ACE/SIS complete two solar cycle measurements.

(2) At energies from 7.3 MeV nuc$^{-1}$ to $E_t$, the ACR-dominated spectra are negatively indexed with $\gamma_1$ varying from -2.0 to -0.5, whereas the GCR-dominated spectra are positively indexed with $\gamma_2$ ranging from 0.3 to 0.8 at energies $E_t$ to 237.9 MeV nuc$^{-1}$. Both $\gamma_1$ and $\gamma_2$ exhibit a positive correlation with SSN, reflecting that the ACR-dominated spectrum gets softer at solar minimum but harder at solar maximum, whereas the GCR-dominated spectrum becomes harder at solar minimum but softer at solar maximum.

(3) During the extremely quiet solar minimum $P_{24/25}$, the peak value of GCR intensities reaches the highest level since the space age, while the maximum ACR intensity is roughly the same as the previous measurements during the $A >0$ solar minina $P_{20/21}$ and $P_{22/23}$. This indicates that the modulation process of ACR and GCR is different. By comparing with HCS tilt angle, we find that ACR intensity is more sensitive to the change of tilt angle, indicating that current sheet drift is more important to ACR component. Detailed numerical simulation is particularly in need for fully understanding different ACR and GCR modulations.

\acknowledgments
This work is supported by the Science and Technology Development Fund, Macau SAR (File Nos. 008/2017/AFJ, 0042/2018/A2, and 0002/2019/APD) and the National Natural Science Foundation of China (Grant No. 11761161001). We acknowledge the ACE CRIS and SIS instrument teams for making the cosmic ray data available at the ACE Science Center (\url{http://www.srl.caltech.edu/ACE/ASC/index.html}). Monthly sunspot numbers are derived from the Solar Influences Data Analysis Center (\url{http://www.sidc.be/}). Neutron monitor count rates are acquired from the Neutron Monitor DataBase event search tool (\url{http://www01.nmdb.eu/}). Heliospheric current sheet tilt angles are derived from the Wilcox Solar Observatory (\url{http://wso.stanford.edu/}). We appreciate beneficial discussions with Dr. Wensai Shang at Shandong University.

\bibliography{my_ref}{}
\bibliographystyle{aasjournal}

\end{document}